\documentclass[11pt,a4paper,english]{elsarticle}
\usepackage{lmodern}
\usepackage[T1]{fontenc}
\usepackage[latin9]{inputenc}
\usepackage{array}
\usepackage{rotfloat}
\usepackage{units}
\usepackage{textcomp}
\usepackage{multirow}
\usepackage{amstext}
\usepackage{graphicx}
\usepackage{setspace}
\makeatletter

\providecommand{\tabularnewline}{\\}

\journal{Materials Chemistry and Physics}

\makeatother

\usepackage{babel}
\begin{document}

\title{Synthesis and structural ordering of nano-sized Ba$_{3}$B$^{'}$Nb$_{2}$O$_{9}$
(B$^{'}$ $=$ Ca and Zn) powders}

\author[uf]{Jo\~ao Elias Figueiredo Soares Rodrigues}
\ead{rodrigues.joaoelias@gmail.com}

\author[uq]{D\'ebora Morais Bezerra}
\author[uq]{Adeilton Pereira Maciel}
\author[uf,bm,bp]{C. W. A. Paschoal\corref{cor1}}
\ead{paschoal@ufma.br; paschoal.william@gmail.com}

\cortext[cor1]{Corresponding author}
\address[uf]{Departamento de F\'{\i}sica, CCET, Universidade Federal do Maranh\~ao, CP
65085-580, S\~ao Lu\'{\i}s, MA, Brazil}
\address[uq]{Departamento de Qu\'{\i}mica, CCET, Universidade Federal do Maranh\~ao,
CP 65085-580, S\~ao Luís, MA, Brazil}
\address[bm]{Department of Materials Science and Engineering, University of California
Berkeley, 94720-1760, Berkeley - CA, United States}
\address[bp]{Department of Physics, University of California Berkeley, 94720-7300,
Berkeley - CA, United States}

\begin{abstract}
In this work we investigated the phase formation
and structural ordering of Ba$_{3}$B$^{'}$Nb$_{2}$O$_{9}$ (B$^{'}$
$=$ Ca and Zn) perovskite powders obtained by a modified polymeric precursor
route using X-ray powder diffraction and Raman spectroscopy. We obtained single phase complex perovskites at 900 \textdegree{}C. Williamson-Hall analysis and scanning
electron microscopy showed that the powders have crystallites and
particles of nanometric size.
\end{abstract}

\maketitle

\section{Introduction}

Ba$_{3}$B$^{'}$B$^{''}$$_{2}$O$_{9}$-type perovskites (B$^{'}$
$=$ Mg, Ca, Zn, Ni, Co, or Cd; B$^{''}$ $=$ Nb and Ta) are materials
widely employed as resonators and filters for wireless communication
technologies due mainly to their high permittivities and low dielectric
losses in the spectral range of micro- and millimeter-waves \cite{Sebastian2008,Sagala-Nambu1992,SinyBhalla1998,Desu1985,Tamura1986,Tamura1991,Hughes2001,Tsai2003,Dias-Paschoal2003,Lin2007}.
Some of these materials are also applied as non lead based ferroelectric/relaxor
ceramics \cite{Pastor2013,Zhao1990,Promsawat2013549,Wu201287}, high temperature proton conductors
for fuel cells \cite{Valdez-Ramirez2012,Ruiz-Trejo2005}, and photocatalysts
which split water into H$_{2}$ and O$_{2}$ under ultraviolet light
irradiation \cite{Xu2007,Yin2004a}.

These oxides belong to the family of complex perovskites, being classified
into disordered and ordered, depending on the degree of ordering of
B$^{'}$ and B$^{''}$ ions. The disordered structure adopts a cubic
unit cell with space group $Pm\bar{3}m$ (no. 221), in which B$^{'}$
and B$^{''}$ ions are randomly distributed. Otherwise, the ordered
structure has a trigonal unit cell that belongs to the space group
$P\bar{3}m1$ (no. 164) induced by distortion along the $\bigl\langle111\bigr\rangle$
direction of the cubic cell \cite{Galasso1969,Galasso-Pyle1963}.

In recent years, several works demonstrated that the structural ordering
drives the physical and chemical properties of Ba$_{3}$B$^{'}$B$^{''}$$_{2}$O$_{9}$-type
perovskites. For example, the features of optical polar phonons and
dipole interactions carry on the high permittivity of ordered
Ba$_{3}$ZnNb$_{2}$O$_{9}$, being very sensitive to the local disorder
\cite{Kamba2004,Lee2007a}. Meanwhile, the protonic conductors based
on the non stoichiometric series Ba$_{3}$Ca$_{1+x}$Nb$_{2-x}$O$_{9-\delta}$
proved that their reductions are strongly influenced by calcium excess,
and consequently by the disorder \cite{Valdez-Ramirez2012}. Ordered
ceramics are usually prepared by high temperature solid state reaction
of the starting oxides, being required long steps of heat treatment.
Therefore, there is a need for new routes that would enable synthesis
of both ordered and disordered materials at relatively low temperatures,
mainly of applied materials as devices that require low co-firing
temperatures. %Furthermore, the structural ordering at long-range of
%Ba$_{3}$B$^{'}$B$^{''}$$_{2}$O$_{9}$-type perovskites requires
%to be investigated for the purpose of to discover new materials and
%improve the existing ones.

Raman spectroscopy is a powerfull tool to probe the ordering features
in complex perovskites based on their phonon properties, as described in Refs. \cite{Siny1998,Dias2001,RatheeshA2000,Janaswamy2008,Grebennikov2010},. For example,
the crystal structures and phonons of (Ba$_{1-x}$Sr$_{x}$)(Mg$_{\nicefrac{1}{3}}$Nb$_{\nicefrac{2}{3}}$)O$_{3}$
solid solutions were studied using this technique together X-ray powder
diffraction, probing a phase transition for samples with strontium
excess \cite{Dias2007}. Also, the evolution of the structural ordering
in Ba$_{3}$CaNb$_{2}$O$_{9}$ ceramics were well probed by Raman
spectroscopy \cite{Rodrigues2013}. Usually,
the vibrational spectra of complex perovskites are predicted in agreement
with group-theoretical formalism, where there is a very close correlation
between spectroscopic and ordering features \cite{Moreira2001,Wang2009,Diao2012,Liu1998}.

This work investigates the structural ordering of Ba$_{3}$CaNb$_{2}$O$_{9}$
(BCN) and Ba$_{3}$ZnNb$_{2}$O$_{9}$ (BZN) complex perovskites using
powders prepared by a modified polymeric precursor route. We probed
the structural features using the X-ray powder diffraction associated
to the Rietveld refinement analysis, scanning electron microscopy,
and Raman spectroscopy. Our main aims were to evaluate the single phase formation and structural ordering
induced by the calcining process. %Finally, we hope that this work
%yields some relevant ideas to future studies concerning the ordering
%process in ceramic materials.

\section{Material and methods}

\subsection{Chemicals and materials}

The compounds barium nitrate (Ba(NO$_{3}$)$_{2}$, Sigma-Aldrich Co.),
calcium citrate tetrahydrate (Ca$_{3}$(C$_{6}$H$_{5}$O$_{7}$)$_{2}$$\cdot$4H$_{2}$O,
Ecibra), zinc nitrate (Zn(NO$_{3}$)$_{2}$, Sigma-Aldrich Co.), ammonium
complex of niobium (NH$_{4}$H$_{2}${[}NbO(C$_{2}$O$_{4}$)$_{3}${]}$\cdot$3H$_{2}$O,
CBMM), ethylene glycol (C$_{2}$H$_{6}$O$_{2}$, Merck), and citric
acid (C$_{6}$H$_{8}$O$_{7}$$\cdot$H$_{2}$O, Proquímico) were used as starting
materials to obtain polymeric precursors. All the reagents had pa grade and they were used as acquired. Distilled water was used throughout the experiments and the heat treatments were carried out in a conventional furnace.

\subsection{Preparation of powders}

The samples were prepared using a modified polymeric precursor (MPP) route
based on the Pechini method \cite{Pechini67}. The precursors based
on barium and zinc  were obtained in accordance with
the following steps. First, the aqueous solutions of nitrate and citric
acid were mixed in a molar ratio of 1:3 metal-citric acid, and kept
under stirring between 60 and 70 \textdegree{}C. Then, ethylene glycol
was added to metal-citric acid aqueous solution in a mass ratio of
1:1 in relation to citric acid. The procedures used to prepare the calcium
polymeric precursors were similar to that employed for the barium and zinc precursors. To obtain niobium polymeric precursor, niobium hydroxide (Nb(OH)$_{5}$) was precipitated by stirring an aqueous solution of ammonium complex of niobium until pH $\approx$ 9 in a thermal bath at 0 \textdegree{}C. Niobium hydroxide was separated from oxalate ions using vacuum filtering and then it was washed with water at 40 \textdegree{}C. Thus, citric acid and ethylene glycol were added to the niobium hydroxide
solution.

In the mixing process of the polymeric precursors, the precursors were kept at the same pH value to avoid precipitation. To determine the amount of each precursor required to obtain the correct chemistry stoichiometry, gravimetric analysis at 900 \textdegree{}C for 1h were employed. After, the precursors of barium, calcium (zinc),
and niobium were mixed to produce the polymeric precursor of BCN (BZN). After, these precursors were heated between 80 and 90 \textdegree{}C to form a polyester, which had high viscosity and glassy aspect. The polyesters were treated thermally at 400 \textdegree{}C for 2 h to obtain black porous powders. These powders were grounded using an agate mortar. Finally, the grounded powders were calcined at 500, 700, 900, 1100, and 1300 \textdegree{}C for 2h in covered alumina crucibles to obtain BCN and BZN samples.

\subsection{Characterization of powders}

The crystalline structure of BCN and BZN powders were characterized by X-ray powder
diffraction (XRPD) using a Brucker D8 Advance with CuK$\alpha$ radiation
($\lambda=0.15406$ nm) in the $2\theta$ range between 10 and 100\textdegree{}
with a step size of 0.02\textdegree{} and a total exposure time of
10 h. The XRPD patterns were compared with data from Inorganic Crystal
Structure Database (ICSD, FIZ Kalrsruhe and NIST). The powder morphology and particle size were observed and
estimated using a scanning electron microscopy (SEM$-$Jeol 6360LV) operated at 20 kV of an accelerating
voltage. The Raman spectroscopy measurements were recorded in the wavenumber range between 120 and 1000 cm$^{-1}$ using an iHR550 Horiba scientific spectrometer coupled to an Olympus microscope model BX-41. A He-Ne laser (632.8 nm, 10 mW) was used to excite the Raman signal of the samples. The backscattering geometry was employed to acquire the scattered light, which were collected in an air-cooled Synapse CCD detector. The spectral resolution was kept lower than 2 cm$^{-1}$ using an 1800 grooves/mm
grating in the spectrometer. All measurements were performed at room temperature.

\subsection{Data analysis}

The DBWS-9807 free software was employed to refine the BCN and BZN
structures. The Rietveld refinement method was used to calculate the structural parameters using the Pseudo-Voigt
function in the refinement for profile fitting \cite{Rietveld1967,Rietveld1969,Young1995,Bleicher2000}.
The crystallite size ($D$) and microstrain ($\epsilon$) were calculated
using the Williamson-Hall (HW) analysis \cite{Williamson1953}, in
which both parameters are deconvoluted in the full width at half maximum
(FWHM, $\beta$) of the diffraction peak according to the following
equation
\begin{equation}
\frac{\beta\cos\theta}{\lambda}=\frac{\mathsf{k}}{D}+\frac{4\epsilon}{\lambda}\sin\theta,
\end{equation}
where $\lambda$ is the wavelength of incident radiation, $\mathsf{k}$
is a dimensionless shape factor with typical value of about 0.9, and
$\beta$ is corrected for the instrumental broadening, as follows
\begin{equation}
\beta=\sqrt{\beta_{\mathsf{expt}}^{2}-\beta_{\mathsf{inst}}^{2}}.
\end{equation}
In this method, $\beta_{\mathsf{expt}}$ is the measured broadening and $\beta_{\mathsf{inst}}$
is the instrumental broadening using the Cagglioti parameters
($\mathrm{U}$,$\mathrm{V}$,$\mathrm{W}$) \cite{Caglioti1958}. Here, $\beta_{\mathsf{inst}}$ was obtained from a plate of sintered corundum (Al$_{2}$O$_{3}$, NIST SRM 1976).
Details on the HW analysis are given elsewhere \cite{KhorsandZak2011,Goncalves2012}.

The deconvolutions of the Raman spectra were performed using Lorentzian
peak functions to determine the wavenumber position ($\omega_{0}$,
cm$^{-1}$) and the full width at half maximum (FWHM, cm$^{-1}$) of
each peak \cite{long2002raman}. All spectra were divided by the
Bose-Einstein thermal factor \cite{hayesscattering}.

\section{Results and discussion}
In our investigation we studied the phase formation, structure, ordering and morphology of the BCN and BZN powder calcined at different temperatures. In next sections we discuss each one of these samples' characteristics.

\subsection{Formation and structural features}
\begin{sidewaysfigure}
\caption{\label{fig:XRPD-patterns}}
\includegraphics{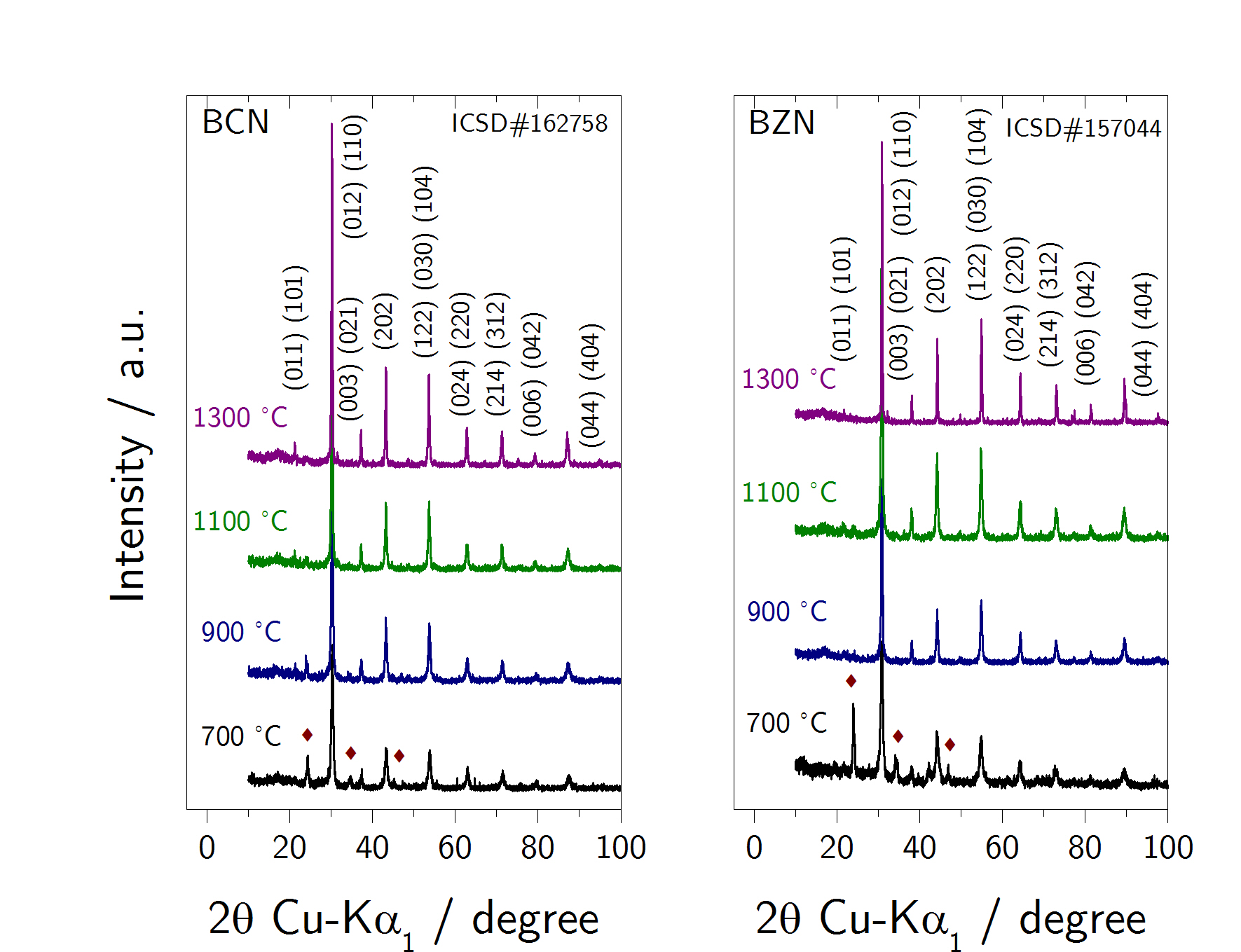}
\end{sidewaysfigure}

The composition and phase purities of the BCN and BZN powders calcined for 2 h at different temperatures were investigated by XRPD. The XRPD patterns obtained are shown in Fig. \ref{fig:XRPD-patterns}. As can be observed, for both BCN and BZN powders, are indexed by the trigonal structure with space group $P\bar{3}m1$ (no. 164, $D_{3d}^{3}$), in agreement with ICSD no. 162758 \cite{Deng2009a,Deng2009b} and ICSD no. 157044 \cite{Lufaso2004,Lufaso2005}, respectively. The refined
parameters for BCN and BZN powders calcined at 1300 \textdegree{}C for 2 h are listed in Table \ref{tab:ref-data}. Diffraction peaks corresponding to the secondary phase of barium niobate phase BaNb$_{2}$O$_{6}$ were clearly observed for powders calcined at 500 and 700 \textdegree{}C. These peaks are marked with diamond symbols in Fig. \ref{fig:XRPD-patterns}. It is outstanding that the single phase was already obtained at 900 \textdegree{}C. This shows the marked availability of our method to obtain good ceramics at low temperatures, mainly because the temperature is lower than 1000 \textdegree{}C, which is wanted to produce low temperature co-fired ceramic (LTCC) devices. Also, it is important to point out that it is hard to prepare materials with complex structures like these perovskites, mainly when MPP is used as synthesis route \cite{Aono2011973}.

\begin{figure}[t]
\caption{\label{fig:HWplot}}
\includegraphics{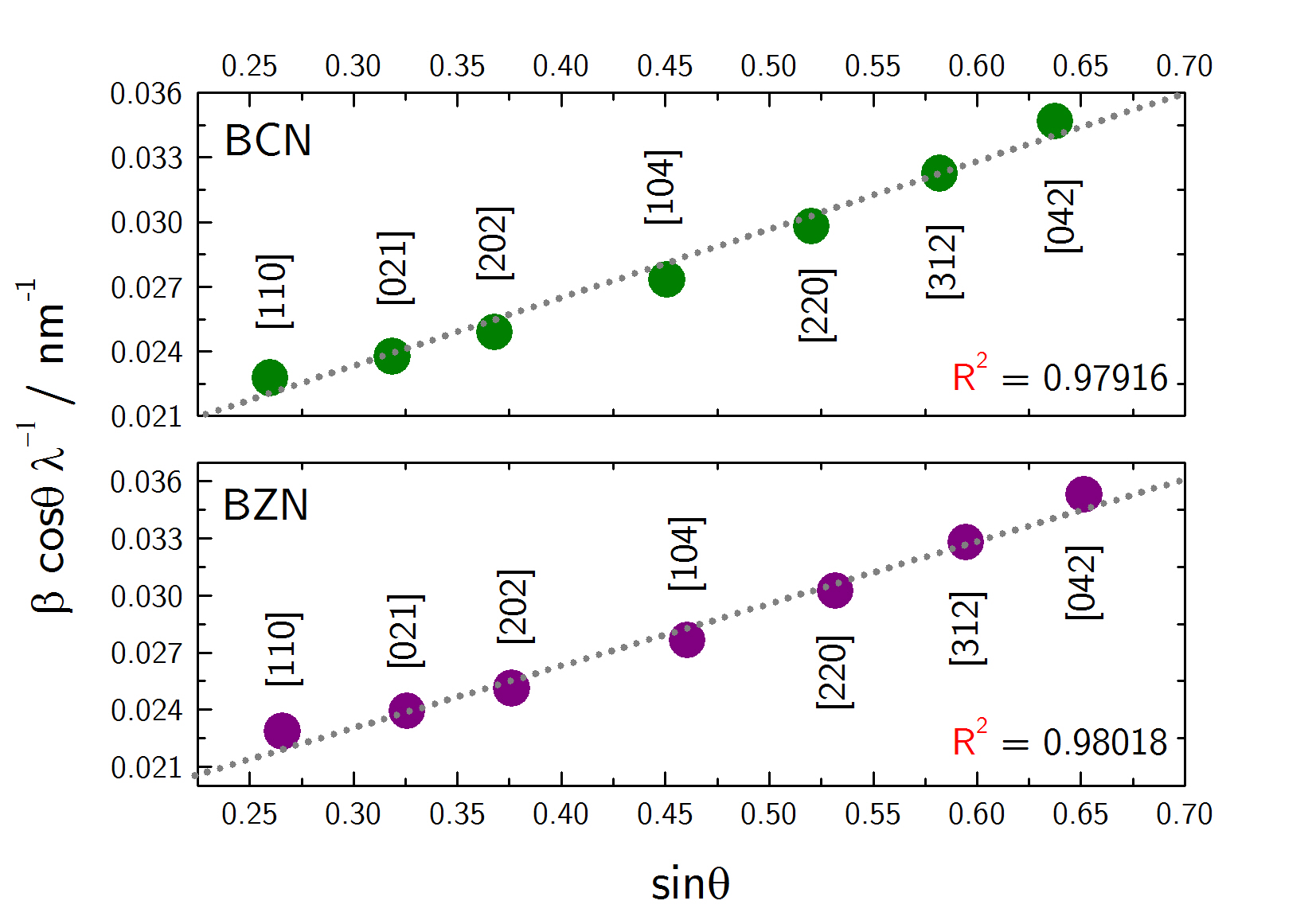}
\end{figure}

MPP usually provides low size crystallites. Thus it is interesting to determine our samples' size and the microstrain, which measures the deformation in the crystallites due to the small size because of surface energies. The crystallite size and microstrain values of the BCN and BZN powders calcined for 2 h at 1300 \textdegree{}C were investigated by WH analyses, which are shown in Fig. \ref{fig:HWplot}. The low fluctuations of the straight lines indicate low dispersion in crystallite size $D$ and microstrain values $\epsilon$. For BCN powders,  $D$ obtained was 65.8 nm and $\epsilon$ was about 0.00123, while $D$ is 67.5 nm and $\epsilon$ was about 0.00126 for BZN powders, showing that both,  BCN and BZN powders calcined at 1300 \textdegree{}C for 2 h, have crystallites of nanometric size.

We also probed the microstrain values using the program STRAIN (available in Bilbao Crystallographic Server). In this program, it is provided two initial unit cells: the undeformed and deformed cells. With basis on these two cells it calculates the linear strain tensor (LST) and its corresponding eigenvalues based on the expansion or contraction of the undeformed structure
with relation to the deformed structure \cite{nye1985physical,haussuhl2007physical}. Here, the undeformed  structures were taken from Refs. \cite{Deng2009b,Lufaso2004} for crystallites with micrometric size, which we considered relaxed. Table \ref{tab:ref-data} summarizes some structural parameters of our undeformed structures and results obtained using the STRAIN program. The positive signal of eigenvalue $\epsilon_{a}$ indicates a lattice expansion in $ab$-plane, while negative signal
of $\epsilon_{c}$ suggests a lattice contraction in $c$-axis of the trigonal unit cell, see Table \ref{tab:williamson-hall}. By comparison with the microstrain values, the eigenvalues of LST show that the effect of lattice strain in BCN and BZN crystallites is larger along $ab$-plane, showing that the partially ordered trigonal structure deforms mainly in $a$- and $b$-axes.

\subsection{Ordering features}

From XRPD patterns it can be estimated the degree of ordering at long-range in materials \cite{cullity1978}.
If the cations B$^{'}$$^{2+}$ and Nb$^{5+}$ are randomly distributed, then the structure is fully disordered. In this case, the pattern should be indexed by the cubic structure that belongs to the space group $Pm\bar{3}m$ (no. 221, $O_{h}^{1}$). Otherwise, if the cations B$^{'}$$^{2+}$ and Nb$^{5+}$ are alternately distributed in the form ( $\cdots$ B$^{'}$$-$Nb$-$Nb$-$B$^{'}$$-$Nb$-$Nb $\cdots$ ) in $\bigl\langle111\bigr\rangle$ direction of the cubic cell, then
the structure is fully ordered and this distribution produces characteristic reflections of the trigonal lattice that belongs to the space group $Pm\bar{3}1$ (no. 164, $D_{3d}^{3}$). \cite{Galasso-Pyle1963}. However, it is important to note that complex perovskites may have different values of degree of ordering, where the lattice assumes a partially ordered trigonal structure, in which B$^{'}$ and Nb ions
can exchange their Wyckoff positions, as described in Refs. \cite{Dias2001,Rodrigues2013,Ioachim2007,Chen2006}.
Nevertheless, the indices of planes in the XRPD pattern can still be indexed by the trigonal structure as we have indexed in Fig. \ref{fig:XRPD-patterns} and showed in previous section. Hence, we assume that BCN and BZN structures have a partial ordering. In addition, our R$\mathsf{_{wp}}$ values are similar for refinements performed using other partially ordered structures, see Refs. \cite{Deng2009b,Deng2007}.

\begin{sidewaysfigure}
\caption{\label{fig:Raman-spectra}}
\includegraphics{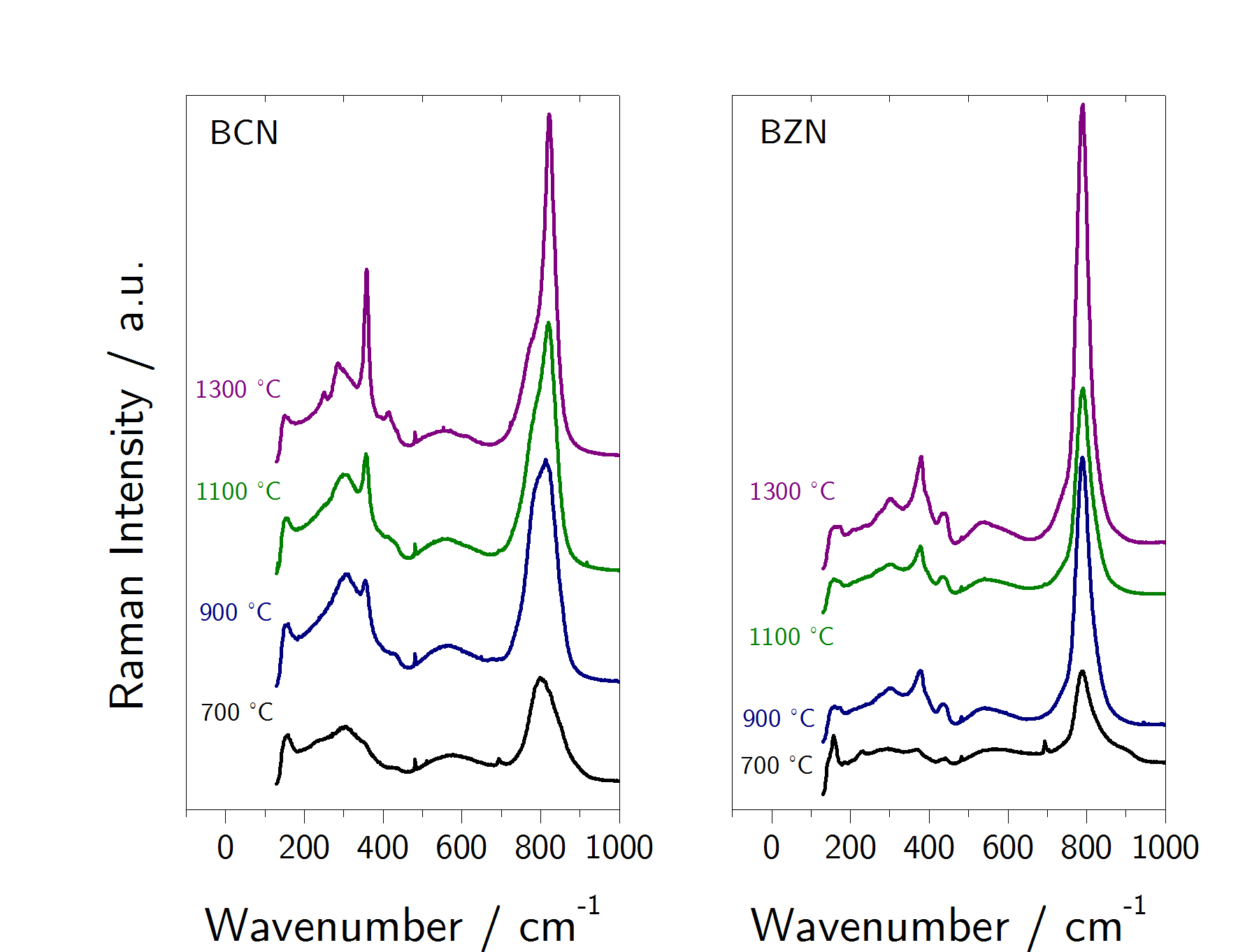}
\end{sidewaysfigure}

As discussed previously, Raman spectroscopy is a powerfull tool to probe the ordering features
in complex perovskites once their phonon properties are based on the site occupation.  Fig. \ref{fig:Raman-spectra} shows the Raman spectra obtained for BCN and BZN powders calcined at different temperatures for 2 h. It is clear that there are significant changes in the Raman spectra of these powders when the calcining temperature is increased. We also note that the Raman bands near 800 cm$^{-1}$ become narrower and more intense. This behavior is associated to the structural ordering, as we discuss in more details next.

According to the group-theoretical prediction, the fully ordered trigonal structure with 15 atoms in its primitive unit cell exhibits 30 phonons, whose distributions are as follows \cite{Deng2009b,Lufaso2004}: one Ba ion occupies 1$a$ site and two others are at 2$d$ sites, giving $\Gamma_{\mathsf{Ba}}=\mathsf{A}_{1g}\oplus\mathsf{E}_{g}\oplus2\mathsf{A}_{2u}\oplus2\mathsf{E}_{u}$;
the B$^{'}$ ion (in our case, B$^{'}$ $=$ Ca, Zn) located at 1$b$ site, for which $\Gamma_{\mathbf{\mathsf{B^{'}}}}=\mathsf{A}_{2u}\oplus\mathsf{E}_{u}$;
two Nb ions are at 2$d$ sites, such that $\Gamma_{\mathsf{Nb}}=\mathsf{A}_{1g}\oplus\mathsf{E}_{g}\oplus\mathsf{A}_{2u}\oplus\mathsf{E}_{u}$;
six O ions are at 6$i$ sites and three others are at 3$e$ sites, giving $\Gamma_{\mathsf{O}}=2\mathsf{A}_{1g}\oplus\mathsf{A}_{2g}\oplus3\mathsf{E}_{g}\oplus2\mathsf{A}_{1u}\oplus4\mathsf{A}_{2u}\oplus6\mathsf{E}_{u}$.
$\mathsf{A}$ modes are non degenerate and $\mathsf{E}$ modes are doubly degenerate, the subscripts $g$ and $u$ indicate the parity property under inversion operation in centrosymmetric crystals, and the symmetry and number of phonons are based on the irreducible representation of the group factor $\bar{3}m$ ($D_{3d}$) \cite{Rousseau1981,BilbaoCS}. After subtracting the silent ($\Gamma_{\mathsf{Silent}}=2\mathsf{A}_{1u}\oplus\mathsf{A}_{2g}$) and acoustic ($\Gamma_{\mathsf{Acoustic}}=\mathsf{A}_{2u}\oplus\mathsf{E}_{u}$) phonons, we expect 9 zone-center Raman-active phonons and 16 zone-center infrared phonons for both BCN and BZN lattices at room temperature, in agreement with the following distributions:
\begin{equation}
\Gamma_{\mathsf{Raman}}=4\mathsf{A}_{1g}\oplus5\mathsf{E}_{g},\label{eq:g-raman}
\end{equation}
\begin{equation}
\Gamma_{\mathsf{IR}}=7\mathsf{A}_{2u}\oplus9\mathsf{E}_{u}.\label{eq:g-ir}
\end{equation}

Otherwise, one must observe an increase in the number of phonons for the partially ordered trigonal structure when there are exchanges in the non equivalent positions of B$^{'}$ and Nb ions \cite{SinyBhalla1998,Tamura1986,Moreira2001,Wang2009,Dias2003}. Indeed, the existence of B$^{'}$ ions at 2$d$ sites provides more 2 Raman-active and 2 IR-active phonons, while Nb ion at 1$b$ site adds just 2 IR-active phonons, as summarized below
\begin{equation}
\Gamma_{\mathsf{B^{'}\,[2d]}}=\mathsf{A}_{1g}\oplus\mathsf{E}_{g}\oplus\mathsf{A}_{2u}\oplus\mathsf{E}_{u},\label{eq:bat2d}
\end{equation}
\begin{equation}
\Gamma_{\mathsf{Nb\,[1b]}}=\mathsf{A}_{2u}\oplus\mathsf{E}_{u}.\label{eq:nbat1b}
\end{equation}
Hence, in partially ordered trigonal structure are expected 11 Raman- and 20 infrared-active phonons in the vibrational spectra. We employed this model to investigate the structural ordering process under different calcining temperatures in BCN and BZN powders.

\begin{sidewaysfigure}
\caption{\label{fig:Raman-decon}}
\includegraphics{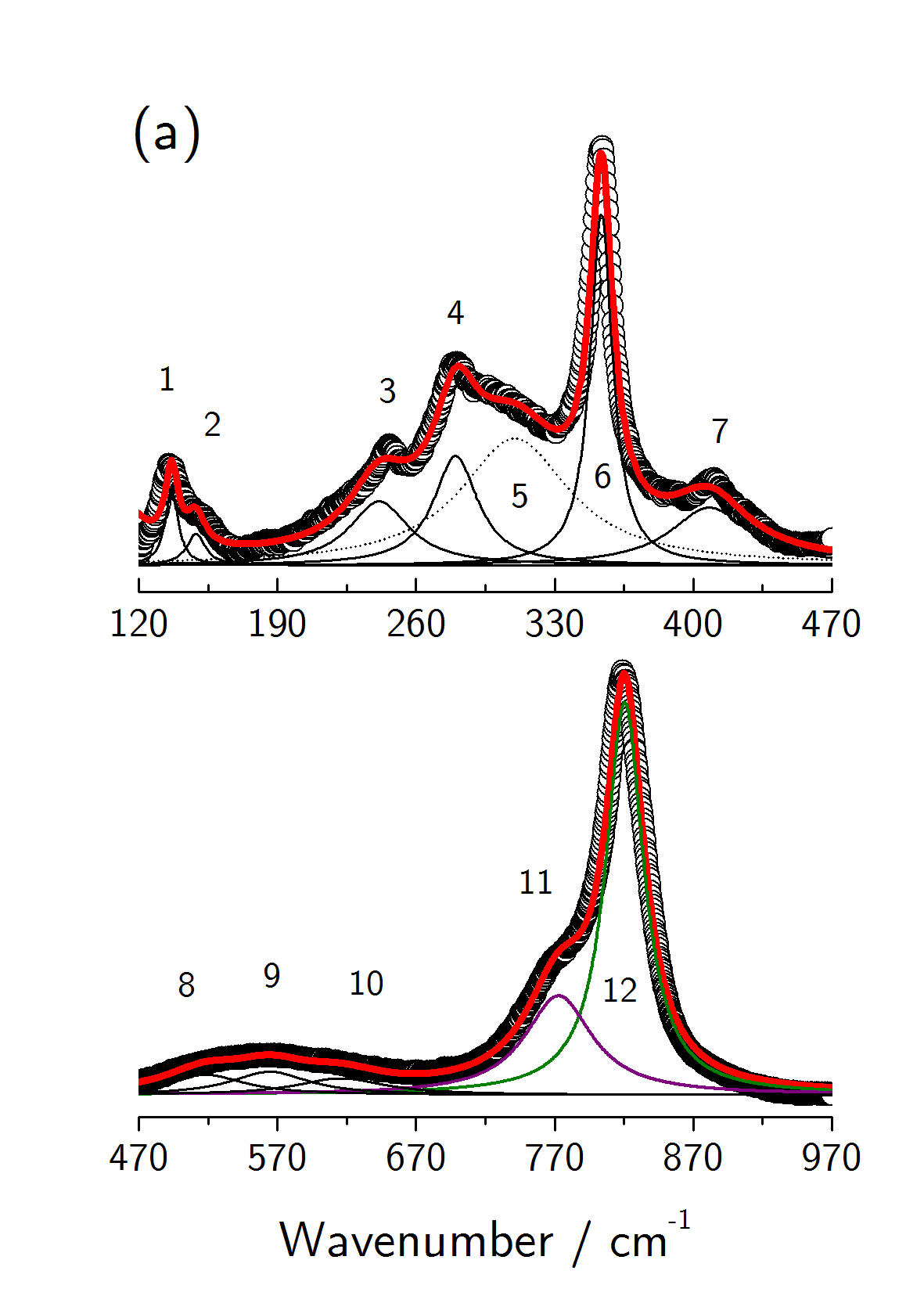}\includegraphics{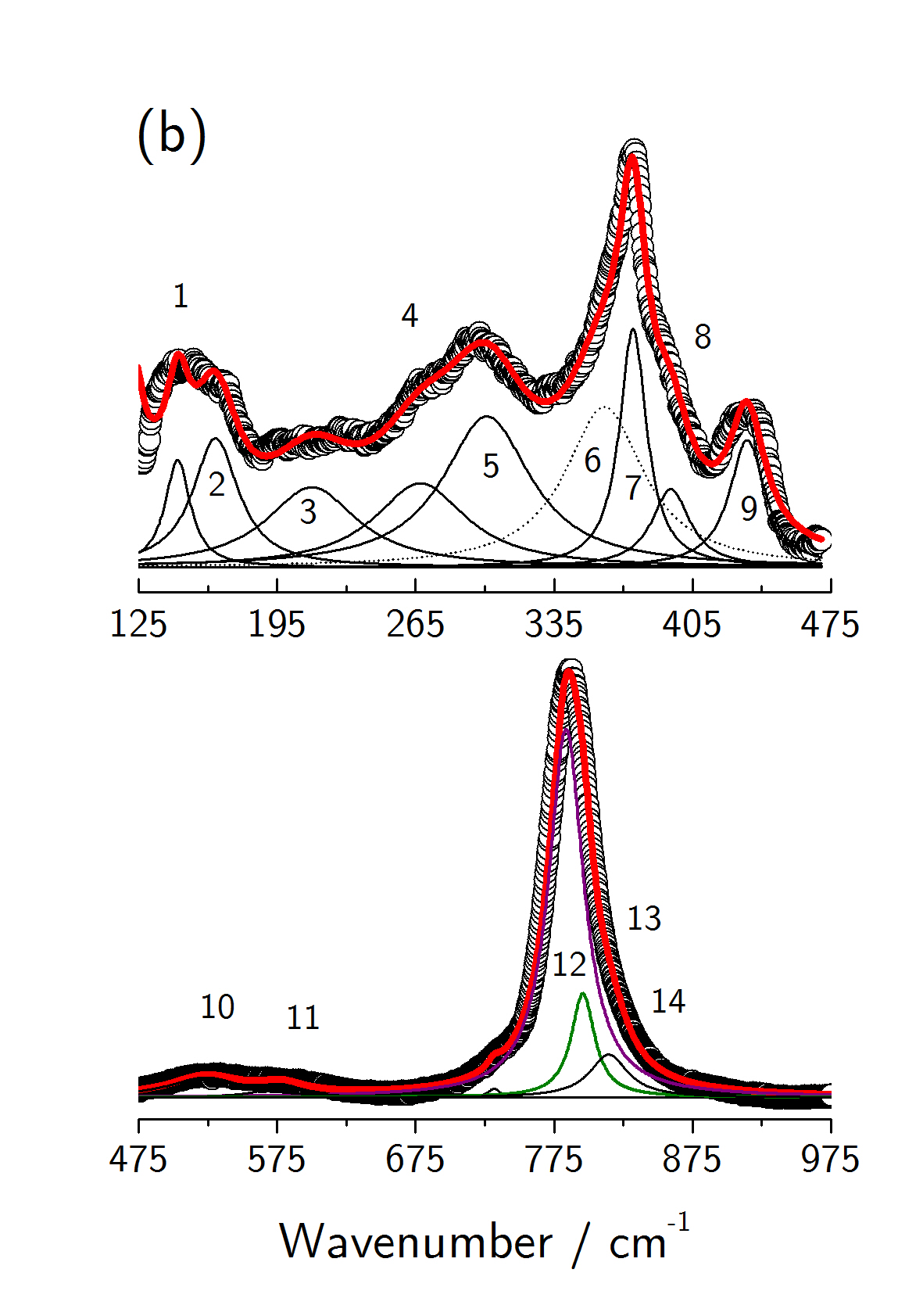}
\end{sidewaysfigure}

From the deconvolution showed in Fig. \ref{fig:Raman-decon}, it was observed 12 and 14 peaks in the Raman spectra of BCN and BZN powders calcined at 1300 \textdegree{}C for 2 h. The measured spectra of BCN and BZN was compared with the Raman-active phonons calculated in Refs. \cite{Rodrigues2013,Dai2009}, as summarized in Table \ref{tab:assignment}. There is an excellent
agreement between the measured and calculated values for the symmetric stretching mode of BCN near 821 cm$^{-1}$. Furthermore, the difference in wavenumber between the measured and calculated data is no more than 25 cm$^{-1}$ for each mode of BZN. The phonons that are directly related with the fully ordered trigonal structure are recognized in the range between 130 and 305 cm$^{-1}$ (see peak nos. 1, 3, and 4 of BCN and peak nos. 2, 4 and 5 of BZN). Also, it can be observed the extra phonons in the Raman spectra of BCN and BZN at, respectively, 615.9 and 523.7 cm$^{-1}$, which are also connected with the fully ordered
structure \cite{Wang2009,Dai2009,Ning2012}. Besides, some peaks attributed to defects in our samples (DM), and baseline effect in the deconvolution process (FLB) were identified \cite{Moreira2001,Wang2009}. These defect modes are usual in ceramic materials \cite{Prosandeev2005}.

\begin{sidewaysfigure}
\caption{\label{fig:Raman-800}}
\includegraphics{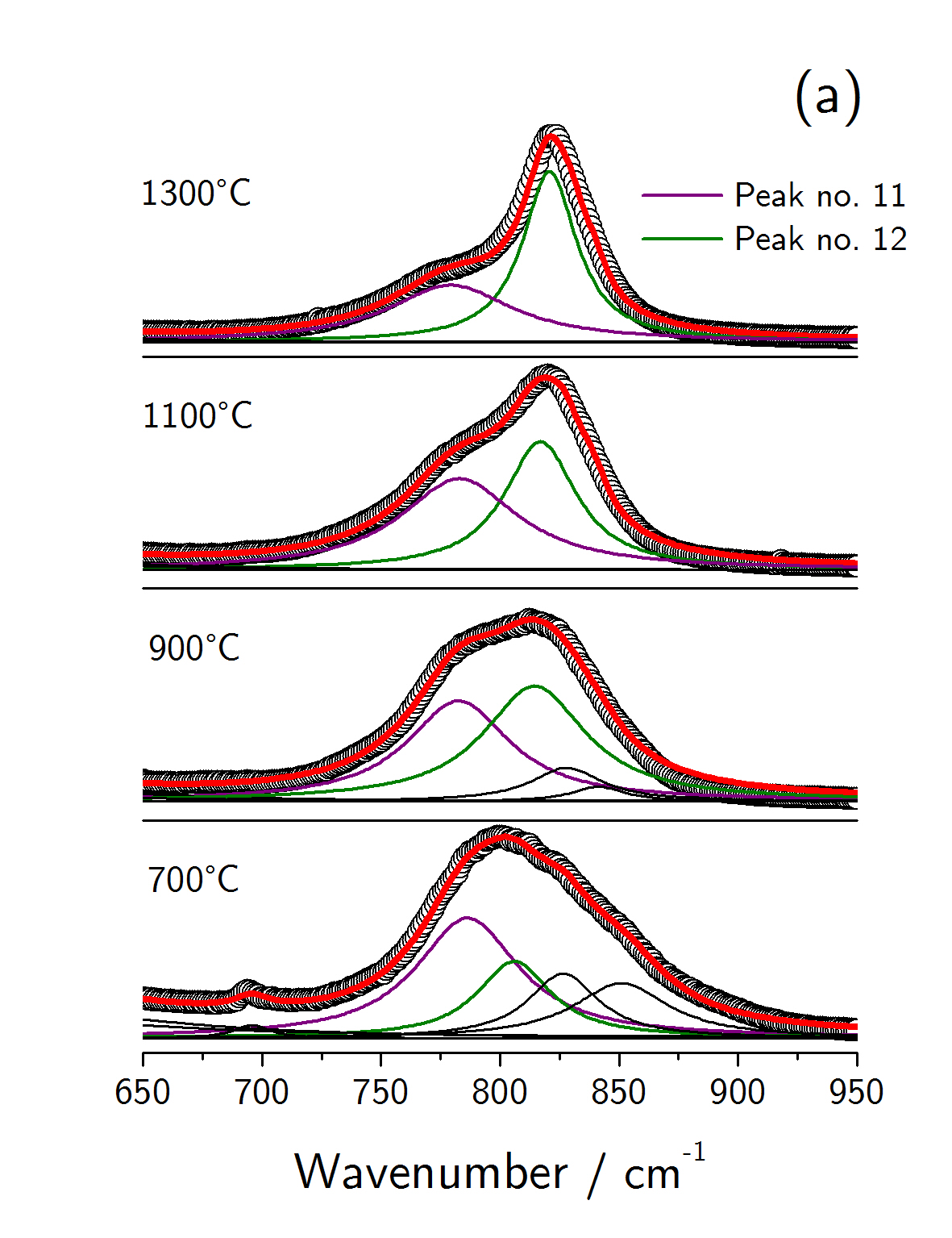}\includegraphics{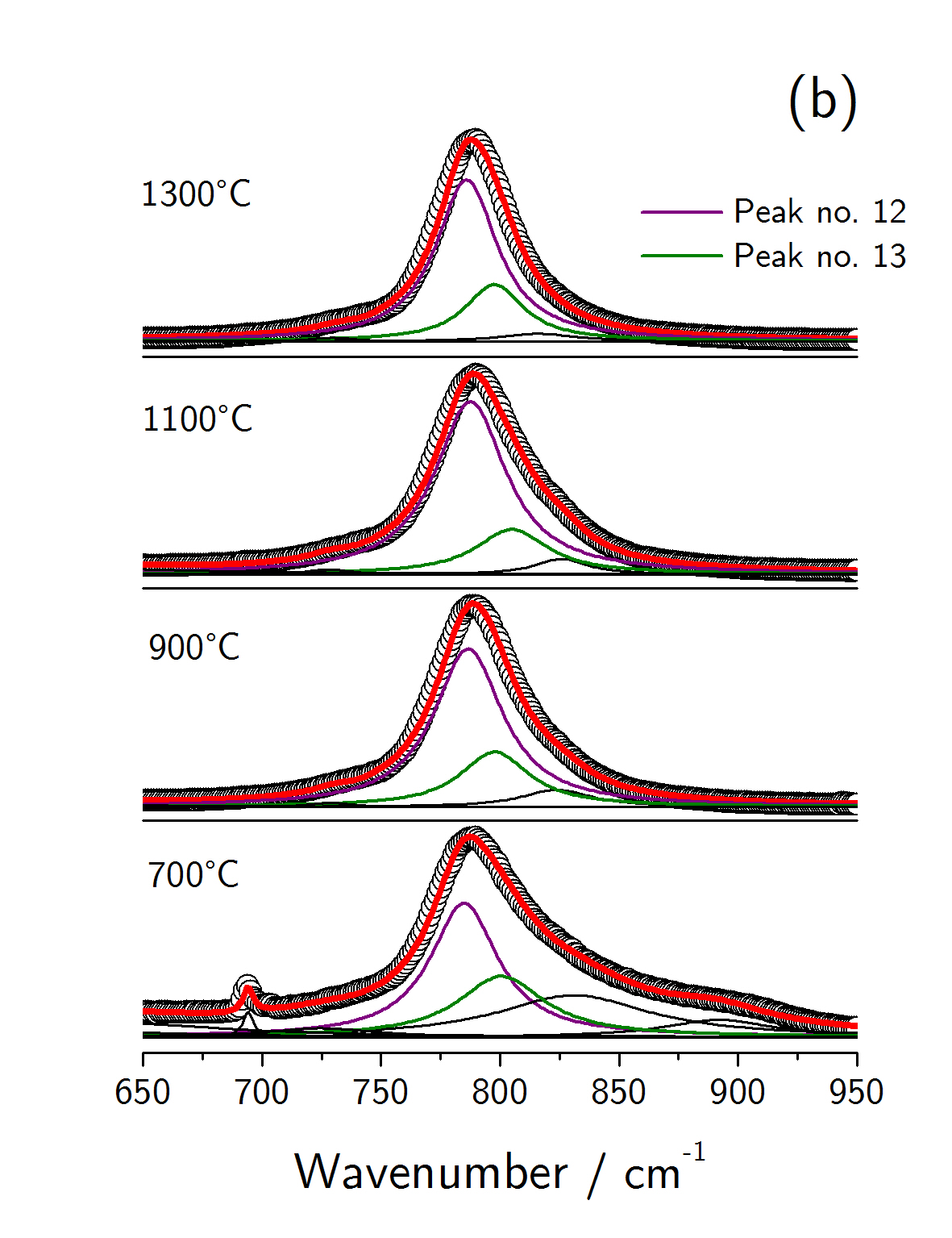}
\end{sidewaysfigure}

Since the modes $\mathsf{A}_{1g}^{(3)}$ and $\mathsf{A}_{1g}^{(4)}$ are related to breath-vibrations of oxygen octahedron, and the modes $\mathsf{E}_{1g}^{(3)}$ and $\mathsf{E}_{1g}^{(4)}$ are connected to twisting breath-vibrations of oxygen octahedron \cite{Dias2007,Wang2009,Ning2012}, the vibrations of oxygens at 6$i$ sites are strongly influenced by
B$^{'}$ ions occupation at 2$d$ sites in the partially ordered structure due to the different strength between Ca$-$O bond, Zn$-$O bond and Nb$-$O bond. This behaviour is corroborated through the occurrences of the peak nos. 9 and 11 in the BCN spectra and peak nos. 11 and 13 in the BZN spectra, in agreement with the model described in Eqs. (\ref{eq:bat2d}) and (\ref{eq:nbat1b}). Indeed, the ratios between the wavenumbers of peak nos. 9$-$10 (0.92) and 11$-$12 (0.94) are,
respectively, 4.3$\%$ and 2.1$\%$ less than the estimated value (0.96) based on the charge-mass relationship $\left[q\,_{\mathsf{Ca}}/m\,_{\mathsf{Ca}}\right]^{\frac{1}{2}}/\left[q\,_{\mathsf{Nb}}/m\,_{\mathsf{Nb}}\right]^{\frac{1}{2}}$
\cite{Moreira2001,Castro2009}. Hence, the peak nos. 9 and 11 correspond to the Raman-active phonons of Ca ions at 2$d$ sites, as described in Ref. \cite{Rodrigues2013}. However, the Zn$-$O bond not simply follows the charge-mass relationship. Instead, the dipole interaction should be take account to explain the correct attribution of BZN measured modes due mainly to the strong covalent bond between Zn and O which is related to the zinc 3$d$ orbital \cite{Dai2009,Ning2012,Bellaiche1999}.
In this sense, we further suggest that peak nos. 11 and 13 correspond to the Raman-active phonons of Zn ions at 2$d$ sites. The ratios between the wavenumbers of peak nos. 11$-$10 (1.10) and 13$-$12 (1.01) show that our assumption is correct. Thus, the structural ordering process can be probed by analyzing the evolutions of these measured peaks, as depicted in Fig. \ref{fig:Raman-800}. We observed that some peaks become stronger, and others become weaker in the wavenumber
range between 650 and 950 cm$^{-1}$, when calcining temperature is increased. The peak no. 12 intensifies with increasing calcining temperatures. On the other hand, the intensities of peak nos. 11 (BCN) and 13 (BZN) decrease with increasing calcining temperatures, and thus structural ordering. This behaviour allows us to estimate the degree of ordering
at long-range using Raman spectroscopy \cite{Siny1998,Moreira2001,Blasse1974,Kim1995}. In this sense, we used the ratio of peak intensities designated by $\Psi_{\mathsf{B'(Nb)}}$ and expressed as follows
\begin{equation}
\Psi_{\mathsf{Ca\,(Nb)}}=\frac{\mathrm{I}_{\mathsf{11\,(12)}}}{\mathrm{I}_{\mathsf{11}}+\mathrm{I}_{\mathsf{12}}},
\end{equation}
\begin{equation}
\Psi_{\mathsf{Zn\,(Nb)}}=\frac{\mathrm{I}_{\mathsf{13\,(12)}}}{\mathrm{I}_{\mathsf{13}}+\mathrm{I}_{\mathsf{12}}}.
\end{equation}
If the ratio $\Psi_{\mathsf{Nb}}$ is equal to one, then Nb ions are located at 2$d$ sites and B$^{'}$ ion locates at 1$b$ site. For partially ordered structure, the ratio $\Psi_{\mathsf{B'}}$ is non null in accordance with Eq. (\ref{eq:bat2d}) \cite{Rodrigues2013}. As can be seen in Table \ref{tab:ratio}, there is a progressive increase in ratio $\Psi_{\mathsf{Nb}}$ with increasing the calcining temperatures, showing that modes $\mathsf{A}_{1g}^{(4)}$ are highly sensitive to
change in the structural ordering. Finally, the most ordered nano-sized BCN and BZN powders are that calcined for 2 h at 1300 \textdegree{}C, for which the ratios $\Psi_{\mathsf{Ca}}$ and $\Psi_{\mathsf{Zn}}$ are equal to 0.25 and 0.21, respectively. Also, we can monitor the structural ordering process by monitoring the behaviour of phonons provided from B$^{'}$ ions (in our case, B$^{'}$ $=$ Ca, Zn) at 2$d$ sites.

\begin{figure}[t]
\caption{\label{fig:center-fwhm}}
\includegraphics{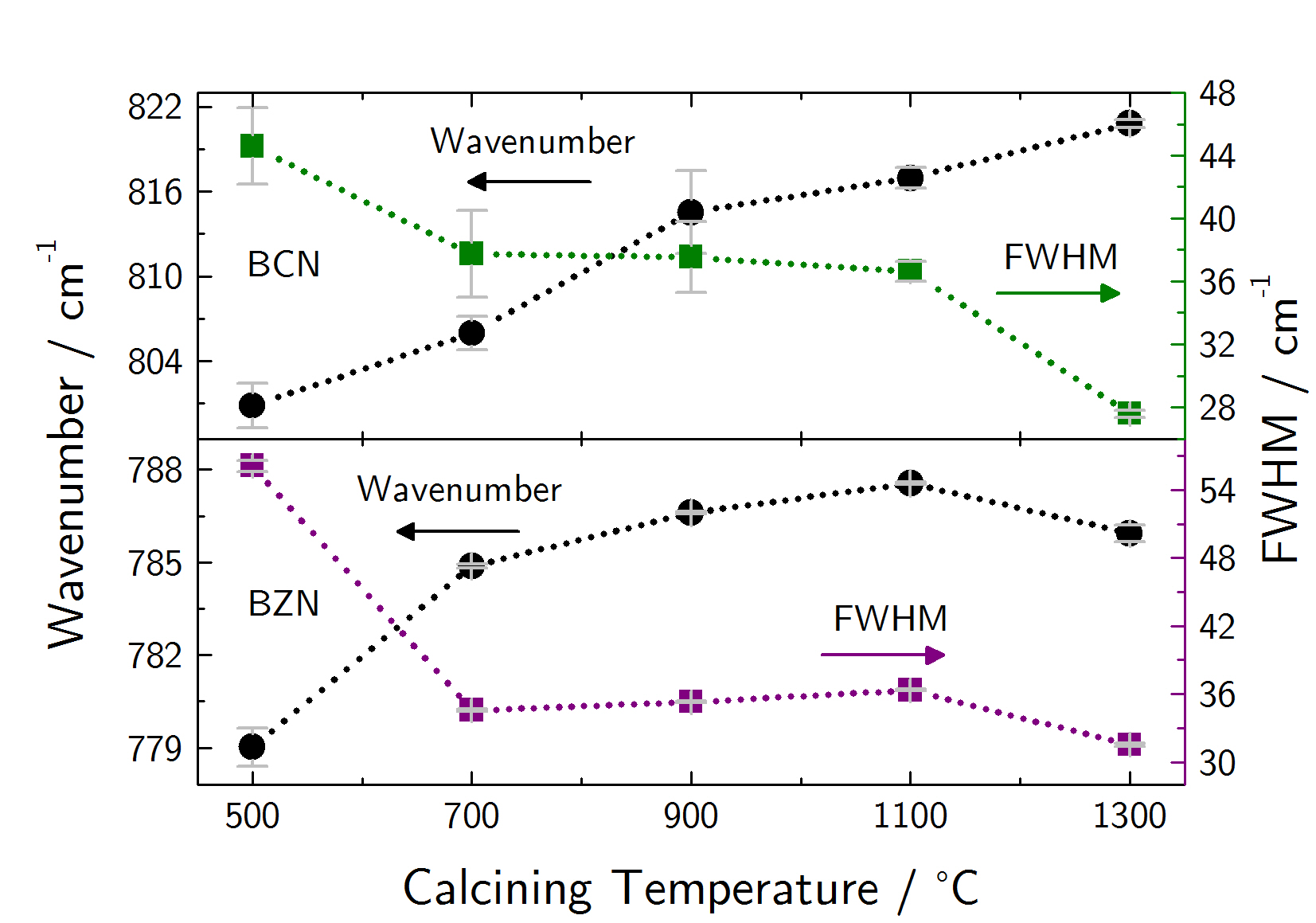}
\end{figure}

The wavenumber position and FWHM calculated from peak deconvolutions confirmed the role of mode $\mathsf{A}_{1g}^{(4)}$
to probe the structural ordering process, and we present these parameters in Fig. \ref{fig:center-fwhm}. Clearly, the wavenumber position of the mode $\mathsf{A}_{1g}^{(4)}$ increases with increasing calcining temperatures, while FWHM decreases with the temperature increase up to 1300 \textdegree{}C. This trend indicates a successive increase in the structural ordering because widest (narrower) peak corresponds to shorter (longer) phonon lifetime and thus more (less) interaction with phonons, in light of the following expression

\begin{equation}
\tau=\frac{1}{\pi c\gamma},
\end{equation}
where $c$ is the velocity of light, and $\tau$ is the phonon lifetime. This equation is based on the energy uncertainty relationship \cite{Anand1996,Bergman1999,Birkedal2001}. Therefore, a narrower FWHM suggests a little damping effect in the
lattice vibrations, and thus a most ordered structure.

\subsection{Morphological features}

\begin{figure}[t]
\caption{\label{fig:SEM-micrographs}}
\includegraphics[scale=0.5]{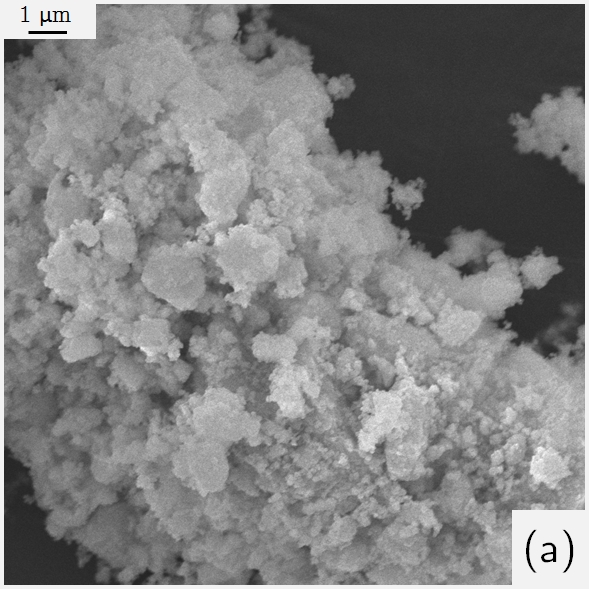}\includegraphics[scale=0.5]{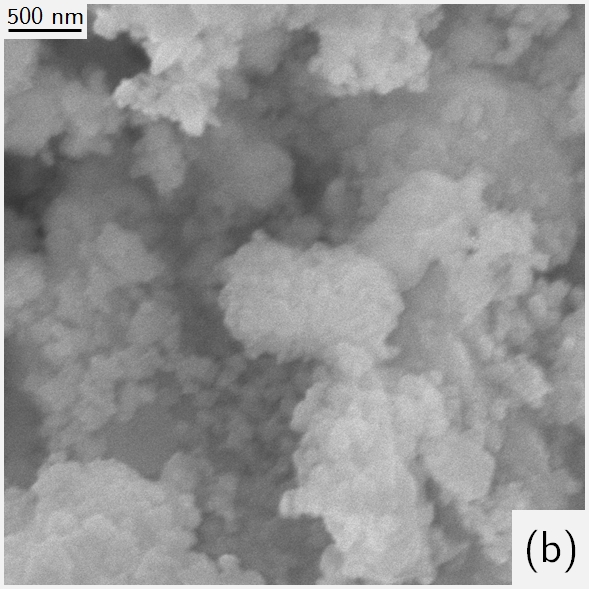}

\includegraphics[scale=0.5]{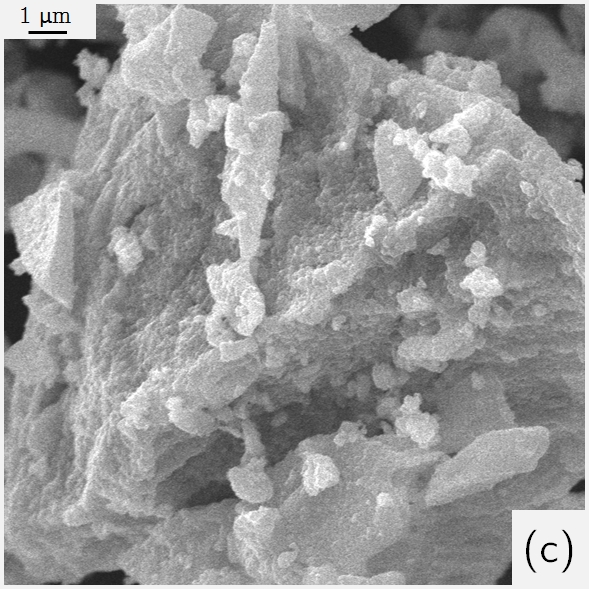}\includegraphics[scale=0.5]{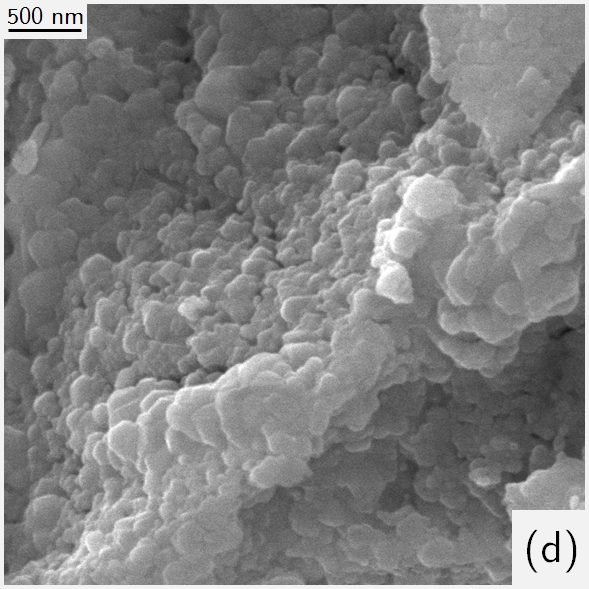}
\end{figure}

Fig. \ref{fig:SEM-micrographs} depicts the SEM micrographs of BCN
and BZN powders calcined for 2 h at 900 \textdegree{}C. SEM micrographs
show that BCN and BZN powders treated thermally at 900 \textdegree{}C
exhibit a large quantity of agglomerated particles. We also observe
the formation of blocks and necks in Figs. \ref{fig:SEM-micrographs}(a)
and \ref{fig:SEM-micrographs}(c) for BCN and BZN, respectively. The
random aggregation process between small particles implied in the
formation of a sphere- and ellipsoid-like morphologies, as shown in
Figs. \ref{fig:SEM-micrographs}(b) and \ref{fig:SEM-micrographs}(d).

\begin{figure}[t]
\caption{\label{fig:The-average-particle}}
\includegraphics{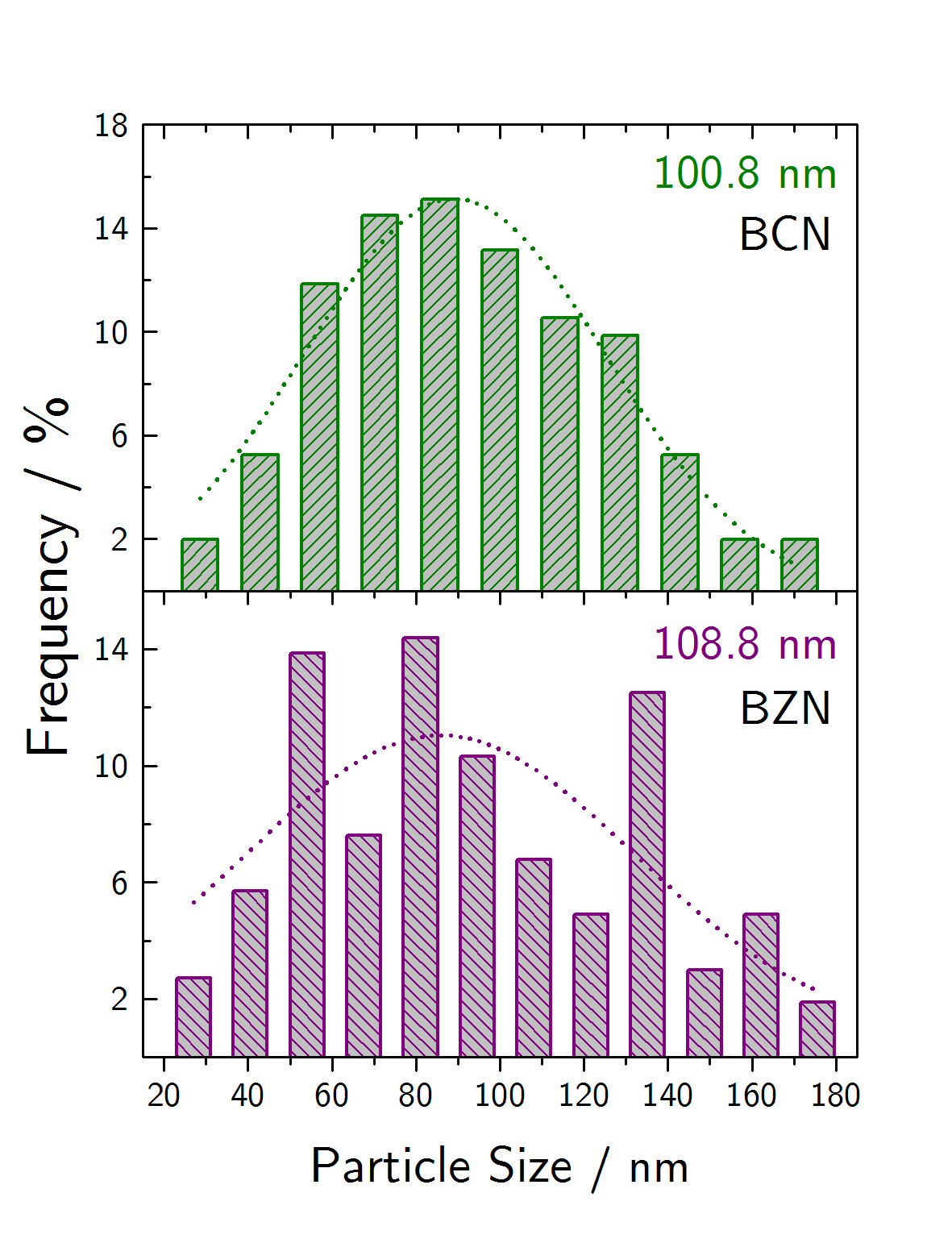}
\end{figure}

The size distributions determined from SEM micrographs depicted in Figs. \ref{fig:SEM-micrographs}(b)
and \ref{fig:SEM-micrographs}(d) are shown in Fig. \ref{fig:The-average-particle}. The average particle sizes of BCN
and BZN powders calcined at 900 \textdegree{}C for 2 h are approximately 100 nm. In addition, the deviations in the particle size distributions can be attributed to the inhomogeneous particles formation due mainly to the synthesis process, which contains a large amount of organic composition that must be obliterated to produce pure oxides. These analyses complement the results discussed from WH analysis.

\section{Conclusions}

A modified polymeric precursor route was used to produce Ba$_{3}$B$^{'}$Nb$_{2}$O$_{9}$
(B$^{'}$ $=$ Ca and Zn) powders. We employed the X-ray powder diffraction and Raman
spectroscopy to evaluate the crystal structure, ordering and phonon behaviours for different calcining temperatures.
It was observed that these polycrystals have a partially ordered structure, which it
belongs to the trigonal space group $P\bar{3}m1$ (no. 164). Single phases were at as low temperatures as 900 \textdegree{}C and the most ordered powders are that calcined for 2h at 1300 \textdegree{}C.
The Williamson-Hall analysis revealed that powders calcined at 1300
\textdegree{}C for 2h had nanometric size in the range of
60$-$70 nm, being deformed mainly in $a$- and $b$-axes of the trigonal
lattice. The SEM micrographs indicated that the average particle sizes
of powders calcined at 900 \textdegree{}C for 2 h are approximately
100 nm, with spherical and ellipsoidal morphologies, besides they showed some inhomogeneity.

\section*{Acknowledgments}

The authors thank the financial support of the Brazilian funding agencies:
CAPES (Coordenação de Aperfeiçoamento de Pessoal de Nível Superior),
CNPq (Conselho Nacional de Desenvolvimento Científico e Tecnológico),
and FAPEMA (Fundação de Amparo à Pesquisa do Estado do Maranhão). C. W. A. Paschoal acknowledges R. Ramesh for all support at Univ. California Berkeley.

%\section*{References}
%
%\bibliographystyle{elsarticle-num}
%\addcontentsline{toc}{section}{\refname}\bibliography{references}

\begin{thebibliography}{10}
\expandafter\ifx\csname url\endcsname\relax
  \def\url#1{\texttt{#1}}\fi
\expandafter\ifx\csname urlprefix\endcsname\relax\def\urlprefix{URL }\fi
\expandafter\ifx\csname href\endcsname\relax
  \def\href#1#2{#2} \def\path#1{#1}\fi

\bibitem{Sebastian2008}
M.~Sebastian, {Dielectric Materials for Wireless Communication}, 1st Edition,
  Elsevier, 2008.

\bibitem{Sagala-Nambu1992}
D.~A. Sagala, S.~Nambu, {Microscopic Calculation of Dielectric Loss at
  Microwave Frequencies for Complex Perovskite Ba(Zn1/3Ta2/3)O3}, Journal of
  the American Ceramic Society 75~(9) (1992) 2573--2575.

\bibitem{SinyBhalla1998}
A.~{Siny, I.G., Tao, R., Katiyar, R.S., Guo, R., Bhalla}, {Raman Spectroscopy
  of Mg-Ta Order-Disorder in BaMg1/3Ta2/3O3}, Journal of Physics and Chemistry
  of Solids 59~(2) (1998) 181--195.

\bibitem{Desu1985}
S.~B. Desu, H.~M. O'Bryan, {Microwave Loss Quality of Ba(Zn1/3Ta2/3)O3
  Ceramics}, Journal of the American Ceramic Society 68~(10) (1985) 546--551.

\bibitem{Tamura1986}
H.~Tamura, D.~A. Sagala, K.~Wakino, {Lattice Vibrations of Ba(Zn1/3Ta2/3)O3
  Crystal with Ordered Perovskite Structure}, Japanese Journal of Applied
  Physics 25~(Part 1, No. 6) (1986) 787--791.

\bibitem{Tamura1991}
H.~Matsumoto, H.~Tamura, K.~Wakino, {Ba(Mg, Ta)O3-BaSnO3 High-Q Dielectric
  Resonator}, Japanese Journal of Applied Physics 30~(Part 1, No. 9B) (1991)
  2347--2349.

\bibitem{Hughes2001}
H.~Hughes, D.~M. Iddles, I.~M. Reaney, {Niobate-based microwave dielectrics
  suitable for third generation mobile phone base stations}, Applied Physics
  Letters 79~(18) (2001) 2952.

\bibitem{Tsai2003}
T.-r. Tsai, C.-c. Chi, M.-h. Liang, C.-T. Hu, I.-N. Lin, {Dielectric properties
  of (x)Ba(Mg1/3Ta2/3)O3-(1-x)Ba(Mg1/3Nb2/3)O3 (x = 1, 0.75, 0.50, 0.25 and 0)
  complex perovskite ceramics}, Materials Chemistry and Physics 79~(2-3) (2003)
  169--174.

\bibitem{Dias-Paschoal2003}
A.~Dias, C.~W.~A. Paschoal, R.~L. Moreira, {Infrared Spectroscopic
  Investigations in Ordered Barium Magnesium Niobate Ceramics}, Journal of the
  American Ceramic Society 86~(11) (2003) 1985--1987.

\bibitem{Lin2007}
I.~N. Lin, C.~T. Chia, H.~L. Liu, H.~F. Cheng, R.~Freer, M.~Barwick, F.~Azough,
  {Intrinsic dielectric and spectroscopic behavior of perovskite
  Ba(Ni1/3Nb2/3)O3-Ba(Zn1/3Nb2/3)O3 microwave dielectric ceramics}, Journal of
  Applied Physics 102~(4) (2007) 044112.

\bibitem{Pastor2013}
M.~Pastor, K.~Biswas, {Synthesis and electrical characterization of
  Ba(Cd1/3Nb2/3)O3 ferroelectric compound}, Materials Chemistry and Physics
  139~(2-3) (2013) 634--639.

\bibitem{Zhao1990}
Z.-L. Zhao, S.-L. Lai, A.-S. Liu, {Ba(Zn1/3Ta2/3)O3-Ba(Cd2/3Nb2/3)O3 dielectric
  ceramics with low microwave loss}, Electronics Letters 26~(19) (1990) 1605.

\bibitem{Promsawat2013549}
M.~Promsawat, J.~Y. Wong, A.~Watcharapasorn, S.~Jiansirisomboon, {Effects of
  sintering temperature on microstructure and electrical properties of
  0.9Pb(Mg1/3Nb2/3)O3-0.1PbTiO3 modified with CuO }, Materials Chemistry and
  Physics 141~(1) (2013) 549 -- 552.

\bibitem{Wu201287}
X.~Wu, L.~Liu, X.~Li, X.~Zhao, D.~Lin, H.~Luo, Y.~Huang, {The influence of
  defects on ferroelectric and pyroelectric properties of
  Pb(Mg1/3Nb2/3)O3-0.28PbTiO3 single crystals}, Materials Chemistry and Physics
  132~(1) (2012) 87 -- 90.

\bibitem{Valdez-Ramirez2012}
O.~Valdez-Ram\'{\i}rez, F.~G\'{o}mez-Garc\'{\i}a, M.~a. Camacho-L\'{o}pez,
  E.~Ruiz-Trejo, {Influence of the calcium excess in the structural and
  spectroscopic properties of the complex perovskite Ba3CaNb2O9}, Journal of
  Electroceramics 28~(4) (2012) 226--232.

\bibitem{Ruiz-Trejo2005}
E.~Ruiz-Trejo, R.~A. {De Souza}, {Dopant substitution and oxygen migration in
  the complex perovskite oxide Ba3CaNb2O9: A computational study}, Journal of
  Solid State Chemistry 178~(6) (2005) 1959--1967.

\bibitem{Xu2007}
B.~Xu, W.~Zhang, X.-Y. Liu, J.~Ye, L.~Shi, X.~Wan, J.~Yin, Z.~Liu,
  {Photophysical properties and electronic structures of the perovskite
  photocatalysts Ba3NiM2O9 (M=Nb,Ta)}, Physical Review B 76~(12) (2007) 125109.

\bibitem{Yin2004a}
J.~Yin, Z.~Zou, J.~Ye, {Photophysical and Photocatalytic Activities of a Novel
  Photocatalyst BaZn1/3Nb2/3O3}, Journal of Physics and Chemistry B 108~(34)
  (2004) 12790--12794.

\bibitem{Galasso1969}
F.~Galasso, {Structure, Properties, and Preparation of Perovskite-type
  Compounds}, 1st Edition, Pergamon Press, 1969.

\bibitem{Galasso-Pyle1963}
J.~{Galasso, F., Pyle}, {Ordering in Compounds of the A(B'0.33Ta0.67)O3 Type},
  Inorganic Chemistry 2~(3) (1963) 482--484.

\bibitem{Kamba2004}
S.~Kamba, H.~Hughes, D.~Noujni, S.~Surendran, R.~C. Pullar, P.~Samoukhina,
  J.~Petzelt, R.~Freer, N.~M. Alford, D.~M. Iddles, {Relationship between
  microwave and lattice vibration properties in Ba(Zn1/3Nb2/3)O3-based
  microwave dielectric ceramics}, Journal of Physics D: Applied Physics 37~(14)
  (2004) 1980--1986.

\bibitem{Lee2007a}
C.-T. Lee, Y.-C. Lin, C.-Y. Huang, C.-Y. Su, C.-L. Hu, {Cation Ordering and
  Dielectric Characteristics in Barium Zinc Niobate}, Journal of the American
  Ceramic Society 90~(2) (2007) 483--489.

\bibitem{Siny1998}
I.~G. Siny, R.~S. Katiyar, A.~S. Bhalla, {Cation arrangement in the complex
  perovskites and vibrational spectra}, Journal of Raman Spectroscopy 29~(5)
  (1998) 385--390.

\bibitem{Dias2001}
A.~Dias, V.~Ciminelli, F.~Matinaga, R.~Moreira, {Raman scattering and X-ray
  diffraction investigations on hydrothermal barium magnesium niobate
  ceramics}, Journal of the European Ceramic Society 21~(15) (2001) 2739--2744.

\bibitem{RatheeshA2000}
R.~Ratheesh, M.~W\"{o}hlecke, B.~Berge, T.~Wahlbrink, H.~Haeuseler,
  E.~R\"{u}hl, R.~Blachnik, P.~Balan, N.~Santha, M.~T. Sebastian, {Raman study
  of the ordering in Sr(B'0.5Nb0.5)O3 compounds}, Journal of Applied Physics
  88~(5) (2000) 2813.

\bibitem{Janaswamy2008}
S.~Janaswamy, G.~S. Murthy, E.~Dias, V.~Murthy, {Structure analysis on the
  Ba3Mg(Ta1-xNbx)O9 ceramics: Coexistence of order and disorder}, Materials
  Research Bulletin 43~(3) (2008) 655--664.

\bibitem{Grebennikov2010}
D.~Grebennikov, O.~Ovchar, A.~Belous, P.~Mascher, {Application of positron
  annihilation and Raman spectroscopies to the study of perovskite type
  materials}, Journal of Applied Physics 108~(11) (2010) 114109.

\bibitem{Dias2007}
A.~Dias, F.~M. Matinaga, R.~L. Moreira, {Raman Spectroscopy of
  (Ba1-xSrx)(Mg1/3Nb2/3)O3 Solid Solutions from Microwave-Hydrothermal
  Powders}, Chemistry of Materials 19~(9) (2007) 2335--2341.

\bibitem{Rodrigues2013}
J.~E. F.~S. Rodrigues, E.~Moreira, D.~M. Bezerra, A.~P. Maciel, C.~W. {de
  Araujo Paschoal}, {Ordering and phonons in Ba3CaNb2O9 complex perovskite},
  Materials Research Bulletin 48~(9) (2013) 3298--3303.

\bibitem{Moreira2001}
R.~L. Moreira, F.~M. Matinaga, A.~Dias, {Raman-spectroscopic evaluation of the
  long-range order in Ba(B'1/3B''2/3)O3 ceramics}, Applied Physics Letters
  78~(4) (2001) 428.

\bibitem{Wang2009}
C.-H. Wang, X.-P. Jing, L.~Wang, J.~Lu, {XRD and Raman Studies on the
  Ordering/Disordering of Ba(Mg1/3Ta2/3)O3}, Journal of the American Ceramic
  Society 92~(7) (2009) 1547--1551.

\bibitem{Diao2012}
C.~Diao, F.~Shi, {Correlation among Dielectric Properties, Vibrational Modes,
  and Crystal Structures in Ba[SnxZn(1-x)/3Nb2(1-x)/3]O3 Solid Solutions}, The
  Journal of Physical Chemistry C 116~(12) (2012) 6852--6858.

\bibitem{Liu1998}
R.~Liu, Y.~Xuan, Y.~Jia, {Ordering and disordering in (A'A'')(B'B'')O3-type
  perovskite compounds}, Materials Chemistry and Physics 57~(1) (1998) 81--85.

\bibitem{Pechini67}
M.~P. Pechini, Method of preparing lead and alkaline earth titanates and
  niobates and coating method using the same to form a capacitor (July 1967).

\bibitem{Rietveld1967}
H.~M. Rietveld, {Line profiles of neutron powder-diffraction peaks for
  structure refinement}, Acta Crystallographica 22~(1) (1967) 151--152.

\bibitem{Rietveld1969}
H.~M. Rietveld, {A profile refinement method for nuclear and magnetic
  structures}, Journal of Applied Crystallography 2~(2) (1969) 65--71.

\bibitem{Young1995}
R.~A. Young, A.~Sakthivel, T.~S. Moss, C.~O. Paiva-Santos, {{\it DBWS}-9411
  {--} an upgrade of the {\it DBWS}*.* programs for Rietveld refinement with PC
  and mainframe computers}, Journal of Applied Crystallography 28~(3) (1995)
  366--367.

\bibitem{Bleicher2000}
L.~Bleicher, J.~M. Sasaki, C.~O. Paiva~Santos, {Development of a graphical
  interface for the Rietveld refinement program {\it DBWS}}, Journal of Applied
  Crystallography 33~(4) (2000) 1189--1189.

\bibitem{Williamson1953}
G.~K. Williamson, W.~H. Hall, {X-ray line broadening from filed aluminium and
  wolfram}, Acta Metallurgica 1~(1) (1953) 22 -- 31.

\bibitem{Caglioti1958}
G.~Caglioti, A.~Paoletti, F.~Ricci, Choice of collimators for a crystal
  spectrometer for neutron diffraction, Nuclear Instruments 3~(4) (1958) 223 --
  228.

\bibitem{KhorsandZak2011}
A.~{Khorsand Zak}, W.~{Abd. Majid}, M.~Abrishami, R.~Yousefi, {X-ray analysis
  of ZnO nanoparticles by Williamson-Hall and size-strain plot methods}, Solid
  State Sciences 13~(1) (2011) 251--256.

\bibitem{Goncalves2012}
N.~Gon\c{c}alves, J.~Carvalho, Z.~Lima, J.~Sasaki, {Size-strain study of NiO
  nanoparticles by X-ray powder diffraction line broadening}, Materials Letters
  72 (2012) 36--38.

\bibitem{long2002raman}
D.~Long, The Raman Effect: A Unified Treatment of the Theory of Raman
  Scattering by Molecules, Wiley, 2002.

\bibitem{hayesscattering}
W.~Hayes, R.~Loudon, Scattering of Light by Crystals, Courier Dover
  Publications, 1978.

\bibitem{Deng2009a}
J.~Deng, J.~Chen, R.~Yu, G.~Liu, X.~Xing, S.~Han, Y.~Liu, D.~Chen, L.~He,
  {Neutron powder diffraction study and B-site ordering in microwave dielectric
  ceramics Ba(Ca1/3Nb2/3)O3}, Solid State Sciences 11~(1) (2009) 170--175.

\bibitem{Deng2009b}
J.~Deng, J.~Chen, R.~Yu, G.~Liu, X.~Xing, {Crystallographic and Raman
  spectroscopic studies of microwave dielectric ceramics Ba(Ca1/3Nb2/3)O3},
  Journal of Alloys and Compounds 472~(1-2) (2009) 502--506.

\bibitem{Lufaso2004}
M.~W. Lufaso, {Crystal Structures, Modeling, and Dielectric Property
  Relationships of 2:1 Ordered Ba3MM'2O9 (M = Mg, Ni, Zn; M' = Nb, Ta)
  Perovskites}, Chemistry of Materials 16~(11) (2004) 2148--2156.

\bibitem{Lufaso2005}
M.~Lufaso, E.~Hopkins, S.~M. Bell, A.~Llobet, {Crystal Chemistry and Microwave
  Dielectric Properties of Ba3MNb2-xSbxO9 (M = Mg, Ni, Zn)}, Chemistry of
  Materials 17~(16) (2005) 4250--4255.

\bibitem{Aono2011973}
H.~Aono, J.~Izumi, M.~Tomida, Y.~Sadaoka, {Synthesis for fine perovskite-type
  materials using precursor prepared by metal nitrates solution mixed with
  organic solvent }, Materials Chemistry and Physics 130~(3) (2011) 973 -- 979.

\bibitem{nye1985physical}
J.~Nye, Physical Properties of Cristals: Their Representation by Tensors and
  Matrices, Oxford University Press, Incorporated, 1985.

\bibitem{haussuhl2007physical}
S.~Hauss\"{u}hl, Physical Properties of Crystals: An Introduction, Wiley, 2007.

\bibitem{cullity1978}
B.~Cullity, Elements of X-Ray Diffraction, Addison-Wesley Publishing Company,
  1978.

\bibitem{Ioachim2007}
A.~Ioachim, M.~Toacsan, M.~Banciu, L.~Nedelcu, C.~Dutu, M.~Feder, C.~Plapcianu,
  F.~Lifei, P.~Nita, {Effect of the sintering temperature on the
  Ba(Zn1/3Ta2/3)O3 dielectric properties}, Journal of the European Ceramic
  Society 27~(2-3) (2007) 1117--1122.

\bibitem{Chen2006}
M.-Y. Chen, C.-T. Chia, I.-N. Lin, L.-J. Lin, C.-W. Ahn, S.~Nahm, {Microwave
  properties of Ba(Mg1/3Ta2/3)O3, Ba(Mg1/3Nb2/3)O3 and Ba(Co1/3Nb2/3)O3
  ceramics revealed by Raman scattering}, Journal of the European Ceramic
  Society 26~(10-11) (2006) 1965--1968.

\bibitem{Deng2007}
J.~Deng, X.~Xing, J.~Chen, R.~Yu, G.~Liu, {Cation ordering in the microwave
  dielectric ceramic BaCd1/3Nb2/3O3}, Scripta Materialia 56~(1) (2007) 65--68.

\bibitem{Rousseau1981}
D.~L. Rousseau, R.~P. Bauman, S.~P.~S. Porto, {Normal mode determination in
  crystals}, Journal of Raman Spectroscopy 10~(1) (1981) 253--290.

\bibitem{BilbaoCS}
E.~Kroumova, M.~Aroyo, J.~Perez-Mato, A.~Kirov, C.~Capillas, S.~Ivantchev,
  H.~Wondratschek, Bilbao crystallographic server: Useful databases and tools
  for phase-transition studies, Phase Transitions 76~(1-2) (2003) 155--170.

\bibitem{Dias2003}
A.~Dias, R.~L. Moreira, {Far-infrared spectroscopy in ordered and disordered
  BaMg1/3Nb2/3O3 microwave ceramics}, Journal of Applied Physics 94~(5) (2003)
  3414.

\bibitem{Dai2009}
Y.~Dai, G.~Zhao, H.~Liu, {First-principles study of the dielectric properties
  of Ba(Zn1/3Nb2/3)O3 and Ba(Mg1/3Nb2/3)O3}, Journal of Applied Physics 105~(3)
  (2009) 034111.

\bibitem{Ning2012}
P.-F. Ning, L.-X. Li, P.~Zhang, W.-S. Xia, {Raman scattering, electronic
  structure and microwave dielectric properties of Ba([Mg1-xZnx]1/3Ta2/3)O3
  ceramics}, Ceramics International 38~(2) (2012) 1391--1398.

\bibitem{Prosandeev2005}
S.~A. Prosandeev, U.~Waghmare, I.~Levin, J.~Maslar, {First-order Raman spectra
  of AB'1/2B''1/2O3 double perovskites}, Phys. Rev. B 71 (2005) 214307.

\bibitem{Castro2009}
M.~C. Castro, E.~F.~V. Carvalho, W.~Paraguassu, A.~P. Ayala, F.~C. Snyder,
  M.~W. Lufaso, C.~W. D.~A. Paschoal, {Temperature-dependent Raman spectra of
  Ba2BiSbO6 ceramics}, Journal of Raman Spectroscopy 40~(9) (2009) 1205--1210.

\bibitem{Bellaiche1999}
L.~Bellaiche, J.~Padilla, D.~Vanderbilt, {Heterovalent and A-atom effects in
  A(B'B'')O3 perovskite alloys}, Physical Review B 59~(3) (1999) 1834--1839.

\bibitem{Blasse1974}
G.~Blasse, A.~Corsmit, {Vibrational spectra of 1:2 ordered perovskites},
  Journal of Solid State Chemistry 10~(1) (1974) 39--45.

\bibitem{Kim1995}
K.~S. {Kim, B.K., Hamaguchi, H., Kim, I. T, Hong}, {Probing of 1-2 Ordering in
  Ba(Ni0.33Nb0.66)O3 and Ba(Zn0.33Nb0.66)O3 ceramics by XRD and Raman
  Spectroscopy}, Journal of the American Ceramic Society 78~(11) (1995)
  3117--3120.

\bibitem{Anand1996}
S.~Anand, P.~Verma, K.~Jain, S.~Abbi, {Temperature dependence of optical phonon
  lifetimes in ZnSe}, Physica B: Condensed Matter 226~(4) (1996) 331--337.

\bibitem{Bergman1999}
L.~Bergman, D.~Alexson, P.~L. Murphy, R.~J. Nemanich, M.~Dutta, M.~A. Stroscio,
  C.~Balkas, H.~Shin, R.~F. Davis, {Raman analysis of phonon lifetimes in AlN
  and GaN of wurtzite structure}, Phys. Rev. B 59 (1999) 12977--12982.

\bibitem{Birkedal2001}
D.~Birkedal, K.~Leosson, J.~M. Hvam, Long lived coherence in self-assembled
  quantum dots, Phys. Rev. Lett. 87 (2001) 227401.

\end{thebibliography}

\newpage{}

\section*{Figure captions}

Fig. 1. XRPD patterns of BCN and BZN powders calcined at 700, 900,
1100, and 1300 \textdegree{}C for 2 h. The indices of planes are indexed
in agreement with the trigonal structure that belongs to space group
$P\bar{3}m1$ (no. 164). The brown diamond symbols indicate the barium
niobate phase.

Fig. 2. Williamson-Hall analyses of BCN and BZN powders calcined for
2h at 1300 \textdegree{}C. The dotted lines are the adjusted functions,
see also Table \ref{tab:williamson-hall}.

Fig. 3. Raman spectra of the BCN and BZN powders calcined at 700,
900, 1100, and 1300 \textdegree{}C for 2 h, whose bands near 800 cm$^{-1}$
become more intense and narrower when the calcining temperature is
increased. This fact is associated to the structural ordering process.

Fig. 4. Deconvolutions of the Raman spectra of BCN (a) and BZN (b)
powders calcined for 2 h at 1300 \textdegree{}C, where the number
on each Lorentzian peak function represents the phonons observed,
see also Table \ref{tab:assignment}.

Fig. 5. Deconvolutions of the Raman spectra of BCN (a) and BZN (b)
powders calcined at different temperatures for 2 h only in the wavenumber
range between 650 and 950 cm$^{-1}$.

Fig. 6. Wavenumber position and full width at half maximum of the mode
$\mathsf{A}_{1g}^{(4)}$ in function of the calcining temperature.
The decrease of FWHM is associated to the structural ordering process
when the calcining temperature is increased.

Fig. 7. SEM micrographs of BCN (a)-(b) and BZN (c)-(d) powders calcined
for 2h at 900 \textdegree{}C, in which the particles present morphology
with spherical and ellipsoidal forms. A log-normal distribution describes
the counting of crystallite sizes.

Fig. 8. The particle size distributions of BCN and BZN powders calcined
for 2 h at 900 \textdegree{}C. Both average particle sizes are approximately
100 nm.

\newpage{}
\newpage{}
\newpage{}

\newpage{}
\newpage{}
\newpage{}
\newpage{}
\newpage{}
\newpage{}
\begin{sidewaystable}
\caption{\label{tab:ref-data}Structural parameters of BCN and BZN powders
refined using X-ray powder diffraction data.}
\begin{tabular}{lllll}
\hline
Chemical formula &  & Ba$_{3}$CaNb$_{2}$O$_{9}$ &  & Ba$_{3}$ZnNb$_{2}$O$_{9}$\tabularnewline
Space group; $Z$ &  & $P\bar{3}m1$; 3 &  & $P\bar{3}m1$; 3\tabularnewline
$a$, $\textrm{\AA}$ &  & 5.9154(6) &  & 5.7900(2)\tabularnewline
$c$, $\textrm{\AA}$ &  & 7.2625(9) &  & 7.0967(5) \tabularnewline
$\gamma_{3}$, \textdegree{} &  & 120 &  & 120\tabularnewline
$V$, $\textrm{\AA}^{3}$ &  & 220.090 &  & 206.040 \tabularnewline
Ba$_{2}$ $z$ (2$d$) &  & 0.6760  &  & 0.6743 \tabularnewline
O$_{1}$ $x$ (6$i$) &  & 0.2142  &  & 0.2114 \tabularnewline
O$_{1}$ $z$ (6$i$) &  & 0.3288  &  & 0.3116 \tabularnewline
Density, g/cm$^{3}$ &  & 5.897 &  & 6.345 \tabularnewline
$D$, nm &  & 65.84$\;\pm\;$4.27 &  & 67.47$\;\pm\;$4.56\tabularnewline
$\epsilon\times10^{3}$ &  & 1.2255$\;\pm\;$0.0728  &  & 1.2552$\;\pm\;$0.0726\tabularnewline
Cagglioti parameters ($\mathrm{U}$,$\mathrm{V}$,$\mathrm{W}$) &  & 0.282876, -0.105603, 0.064551 &  & 0.092537, -0.054572, 0.046780\tabularnewline
R$\mathsf{_{p}}$, R$\mathsf{_{wp}}$, R$\mathsf{_{exp}}$, $\%$ &  & 7.37, 9.81, 5.90 &  & 8.37, 10.71, 5.85\tabularnewline
\hline
\end{tabular}
\end{sidewaystable}
\newpage{}
\begin{sidewaystable}
\caption{\label{tab:williamson-hall}Lattice parameters of BCN and BZN cells
considered as undeformed structures and employed in program STRAIN.
The eigenvalues of linear strain tensor (LST) are also presented.}
\begin{tabular}{ccccccccc}
\hline
\multirow{2}{*}{Chemical formula} &  & \multicolumn{4}{c}{Lattice parameters} &  & \multicolumn{2}{c}{Eigenvalues of LST}\tabularnewline
\cline{3-6} \cline{8-9}
 &  & $a$, $\textrm{\AA}$ & $c$, $\textrm{\AA}$ & $\gamma$, \textdegree{} & Ref. &  & $\epsilon_{a}\times10^{3}$ & $\epsilon_{c}\times10^{3}$\tabularnewline
\hline
Ba$_{3}$CaNb$_{2}$O$_{9}$ &  & 5.9037(7) & 7.2636(3) & 120 & \cite{Deng2009b} &  & 1.992 & -0.139\tabularnewline
Ba$_{3}$ZnNb$_{2}$O$_{9}$ &  & 5.78207(6) & 7.09731(11) & 120 & \cite{Lufaso2004} &  & 1.375 & -0.079\tabularnewline
\hline
\end{tabular}
\end{sidewaystable}
\newpage{}
\begin{sidewaystable}
\caption{\label{tab:assignment}Wavenumbers of measured phonons of BCN and
BZN structures with respective attribution. We have also compared
the measured (expt.) and calculated (calc.) data \cite{Rodrigues2013,Dai2009}.
$^{a}$ DM: defect mode and $^{b}$ FLB: floating base line.}

\begin{tabular}{ccccccccc}
\hline
\multicolumn{4}{c}{Ba$_{3}$CaNb$_{2}$O$_{9}$} & \multirow{2}{*}{} & \multicolumn{4}{c}{Ba$_{3}$ZnNb$_{2}$O$_{9}$}\tabularnewline
\cline{1-4} \cline{6-9}
Peak no. & $\omega_{0}$ (expt.) & $\omega_{0}$ (calc.) & Attrib. &  & Peak no. & $\omega_{0}$ (expt.) & $\omega_{0}$ (calc.) & Attrib.\tabularnewline
\hline
 &  & 59.4 & $\mathsf{E}_{g}^{(1)}$ & \multirow{16}{*}{} &  &  & 98.6 & $\mathsf{E}_{g}^{(1)}$\tabularnewline
 &  & 60.4 & $\mathsf{A}_{1g}^{(1)}$ &  &  &  & 100.8 & $\mathsf{A}_{1g}^{(1)}$\tabularnewline
1 & 136.6 & 62.0 & $\mathsf{E}_{g}^{(2)}$ &  & 1 & 144.6 &  & DM$^{a}$\tabularnewline
2 & 148.9 &  & DM$^{a}$ &  & 2 & 163.9 & 160.13 & $\mathsf{E}_{g}^{(2)}$\tabularnewline
3 & 241.5 & 241.2 & $\mathsf{A}_{1g}^{(2)}$ &  & 3 & 212.7 &  & DM$^{a}$\tabularnewline
4 & 280.0 & 250.7 & $\mathsf{E}_{g}^{(3)}$ &  & 4 & 267.5 & 264.7 & $\mathsf{A}_{1g}^{(2)}$\tabularnewline
5 & 309.8 &  & FLB$^{b}$ &  & 5 & 301.1 & 276.9 & $\mathsf{E}_{g}^{(3)}$\tabularnewline
6 & 353.5 & 305.8 & $\mathsf{E}_{g}^{(4)}$ &  & 6 & 360.4 &  & FLB$^{b}$\tabularnewline
7 & 408.3 & 356.8 & $\mathsf{A}_{1g}^{(3)}$ &  & 7 & 374.9 & 360.9 & $\mathsf{E}_{g}^{(4)}$\tabularnewline
8 & 517.4 &  & DM$^{a}$ &  & 8 & 394.0 &  & DM$^{a}$\tabularnewline
9 & 564.6 &  & \multirow{2}{*}{$\mathsf{E}_{g}^{(5)}$} &  & 9 & 432.6 & 409.5 & $\mathsf{A}_{1g}^{(3)}$\tabularnewline
10 & 615.9 & 658.9 &  &  & 10 & 523.7 & 529.9 & \multirow{2}{*}{$\mathsf{E}_{g}^{(5)}$}\tabularnewline
11 & 772.9 &  & \multirow{2}{*}{$\mathsf{A}_{1g}^{(4)}$} &  & 11 & 578.4 &  & \tabularnewline
12 & 820.5 & 827.7 &  &  & 12 & 783.7 & 770.9 & \multirow{2}{*}{$\mathsf{A}_{1g}^{(4)}$}\tabularnewline
 &  &  &  &  & 13 & 795.8 &  & \tabularnewline
 &  &  &  &  & 14 & 814.6 &  & DM$^{a}$\tabularnewline
\hline
\end{tabular}
\end{sidewaystable}
\newpage{}
\begin{sidewaystable}
\caption{\label{tab:ratio}The ratio of peak intensities calculated from BCN
and BZN powders calcined at 700, 900, 1100, and 1300 \textdegree{}C
for 2h.}
\begin{tabular}{ccccccc}
\hline
\multirow{2}{*}{Temperature, \textdegree{}C} &  & \multicolumn{2}{c}{Ba$_{3}$CaNb$_{2}$O$_{9}$} &  & \multicolumn{2}{c}{Ba$_{3}$ZnNb$_{2}$O$_{9}$}\tabularnewline
\cline{3-4} \cline{6-7}
 &  & $\Psi_{\mathsf{Ca}}$ & $\Psi_{\mathsf{Nb}}$ &  & $\Psi_{\mathsf{Zn}}$ & $\Psi_{\mathsf{Nb}}$\tabularnewline
\hline
700 &  & 0.61 & 0.39 &  & 0.31 & 0.69\tabularnewline
900 &  & 0.47 & 0.53 &  & 0.26 & 0.74\tabularnewline
1100 &  & 0.41 & 0.59 &  & 0.22 & 0.78\tabularnewline
1300 &  & 0.25 & 0.75 &  & 0.21 & 0.79\tabularnewline
\hline
\end{tabular}
\end{sidewaystable}

\end{document}